\documentclass[aps,prl,twocolumn,groupedaddress,showpacs]{revtex4}

\usepackage{amsmath}

\newcommand{\mathsym}[1]{{}}

\usepackage[latin1]{inputenc}
\usepackage[english]{babel}
\usepackage{graphicx}
\usepackage{amssymb}
\usepackage{amscd}
\usepackage{rotating}
\usepackage{color}
\usepackage{bm}

\newcommand{\bra}{\begin{array}}
\newcommand{\era}{\end{array}}
\newcommand{\beq}{\begin{equation}}
\newcommand{\eeq}{\end{equation}}
\newcommand{\bqr}{\begin{eqnarray}}
\newcommand{\eqr}{\end{eqnarray}}

\def\BC{\bb C}
\def\_\BC{\bbi C}

\def\( {\left(}
   \def\) {\right)}
\def\[ {\left[}
\def\] {\right]}
\def\no2 {{\textstyle{n\over 2}}}

\newcommand{\lam}{\lambda}

\newcommand{\be}{\beta}

\newcommand{\te}{\theta}

\newcommand{\al}{\alpha}
\newcommand{\app}{\approx}

\newcommand{\da}{\dagger}

\newcommand{\lb}{\label}

\begin{document}

\title{Integer Quantum Hall Effect in Graphene}

\author{\sc Ahmed Jellal}
\affiliation{Saudi Center for Theoretical Physics, Dhahran, Saudi Arabia\\
Theoretical Physics Group,  
Faculty of Sciences, Choua\"ib Doukkali University, 
24000 El Jadida,
Morocco}

\date{\today}

\begin{abstract}

We study the quantum Hall effect in a monolayer
graphene by using an approach based on thermodynamical properties.
This can be done by considering
 a system of {\it Dirac}
particles in an electromagnetic field
and taking into account of the edges effect as a pseudo-potential varying
continuously along the $x$ direction. At low temperature and in the
weak electric field limit, we explicitly determine the thermodynamical
potential. With this, we derive the particle numbers in
terms of the quantized flux and therefore the Hall conductivity  immediately follows.

\end{abstract}

\maketitle

\section{Introduction}

Graphene is a two dimensional (2D) monolayer of graphite
atoms~\cite{gei,neto}. It has a honeycomb
lattice structure of carbon atoms packed in a 2D system. Its single
layer has a band structure analogous to
massless relativistic particle, where the valence and the
conduction bands meet in two in-equivalent points $K$ and $K'$,
 called {\it
  Dirac} points, at the corners of Brillouin zone. The quantum Hall
  effect (QHE)~\cite{Geim, Zheng} in graphene is one of most
remarkable phenomena, not only because of the Hall conductivity is quantized
on plateaus and magnetoconductance vanishing in magnetic field
but also provides a bridge between condensed matter physics and
quantum electrodynamics~\cite{Novo}.

The successful experimental works~\cite{Geim, Zheng} and several theoretical
attempts~\cite{ Gui, Shara, Ben} established the Hall conductivity
expression as 
$\sigma_{\sf H}=4\left(n+{1\over 2}\right){e^2\over h}$
with $n$, an integer including zero, characterizes the
 integer quantum Hall effect (IQHE) in a monolayer
graphene.
The prefactor 4 reflects the two-fold spin and two-fold valley
degeneracy in the graphene band structure. The term $\frac{1}
{2}$ comes from the Berry phase due to
the pseudospin (or valley) precession when a massless (chiral) Dirac particle exercises
cyclotron motion~\cite{Champel}. The conduction
in graphene device
 may be produced by
 two types of charge carriers: the electrons and the holes. The Fermi
 energy changes the position with changing the type of the carriers
 charges~\cite{Novo}, such that this energy is in the valence
 (conduction) band when the holes (electrons) are responsible to
 conduction. The quantization of the Hall conductivity is determined
 also by the fact of the number of the edges states band crossing the
 Fermi level~\cite{fertig}.

Our main objective is to introduce a new approach based on the thermodynamical properties~\cite{jellalTPG}
to study
the quantum Hall effect in
graphene. For this, we consider  {\it Dirac} particles living in a
rectangular plane under the action of the very weak transverse electric
 and a strong  perpendicular magnetic fields. Taking into account of
 a continuum pseudo-potential varying a long of $x$
 axis, we explicitly evaluate the Hall conductivity. As an interesting result,
 we end up with the quantized plateaux characterizing the  integer
 quantum Hall effect in graphene.

This letter is organized as follows. In section 2,
we formulate our problem by setting the Hamiltonian
describing {\it Dirac} particle in the presence of
the electromagnetic fields and involving a pseudo-potential along the $x$-axis.
After some algebra, we diagonalize our Hamiltonian to get
the solutions of the energy spectrum. In section 3, using  {\it Fermi-Dirac} statistics
and Mellin transformation
to explicitly
evaluate the grand thermodynamical potential.
In section 4, we calculate the particle number to end up with the
Hall conductivity and therefore the corresponding filling factors.
Finally, we conclude in last section.

\section{Solutions of the energy spectrum}

We consider a rectangular sheet of graphene parameterized by two sides
$(L_x,L_y)$ and subjected to an electromagnetic field $(\vec E,
\vec B)$. To deal with our task, we describe the present system by the
Hamiltonian
\begin{equation}\lb{1}
H=v_F \vec{\sigma} \vec{\pi}+\sigma_y eEy+\Delta{\tilde p}+g\mu_B\vec{B} \cdot \vec{S}
\end{equation}
where the first term is the Dirac operator in the presence of $\vec
B$ and second is resulting from an applied electric field along
$y$-direction, i.e. $\vec E= E_y \vec e_y$. The continuum
pseudo-potential $\Delta{\tilde p}$ is reflecting the edges effect
contribution and the last one is
the magnetic coupling.
$\vec \sigma$ are Pauli matrices, $g$ is the Land\'e factor,
$v_F\app{c\over 100}$ is the Fermi velocity and $\mu_B$ is the Bohr
magneton.

It is convenient to consider
the Landau
gauge
$\vec A=(-By,0)$
where the momentum operators read as
$
\pi_x=p_x-{eB\over c}y$ and $\pi_y=p_y$.
For simplicity, we decompose~(\ref{1}) into three parts. These are
\beq\lb{4}
H=H_0+ \Delta{\tilde p}+g\mu_B\vec{B} \cdot \vec{S}
\eeq
where $H_0$ is corresponding to the two first terms in \eqref{1}.
This decomposition is helpful in sense that we can treat each part
separately and therefore derive easily the spectrum of~(\ref{4}).

Now solving the eigenvalue equation to end up with
 the eigenvalues
\begin{equation}\label{landaulevels}
E_{n \tilde{p}s}={\rm sgn}(n)\sqrt{2\left(\frac{\hbar v_F}{l_B}\right)^2 |n|} + \Delta{\tilde p} +
  g \mu_B B m_s
\end{equation}
as well as the corresponding eigenfunctions
\begin{equation}\lb{efn}
\Psi_{n\neq 0, k, m_s}={1\over \sqrt{2}}\left(%
\begin{array}{c}
  -{\rm sgn}(n)i\phi_{|n|-1} \\
  \phi_{|n|} \\
\end{array}%
\right) e^{ikx} {\mathcal\Omega}\ \al_{m_s}
\end{equation}
and eigenfunction are
\begin{equation}\lb{herp}
\phi_{n}=
\sqrt{1\over
  2^n \pi^{1/2} n!l_B}
  e^{-{(y-y_0)^2\over 2l_{B}^{2}}}
  H_n\left(y-y_0\over l_B\right)
\end{equation}
where $|n|=0, 1, 2,\cdots$ is
the LL index, $y_0=-k l_{B}^{2}$, the magnetic length $l_{B}=\sqrt{\hbar c\over eB}$  and $H_n$ being the Hermite polynomial.
The zero-energy mode is
\begin{equation}\lb{ef0}
\Psi_{n=0, k, m_s}=\left(%
\begin{array}{c}
  0 \\
  \phi_{0} \\
\end{array}%
\right) e^{ikx} {\mathcal\Omega}\ \al_{m_s}
\end{equation}
where $m_s=\pm{1\over 2}$ is the azimuthal number of spin operator $S_z$ whose
associated
states are
\beq
\al_{1\over 2}=\left(%
\begin{array}{c}
  1 \\
  0 \\
\end{array}%
\right),\qquad  \al_{-{1\over 2}}=\left(%
\begin{array}{c}
  0 \\
  1 \\
\end{array}%
\right)
\eeq
and by convention we choose ${\rm sgn}(0)=0$.

\section{Grand thermodynamical potential}

To achieve our goal we start by determining the grand thermodynamical potential
 (GTP)
 \beq\lb{GTP1}
\Omega=-k_BT \ln\left({\mathcal Z}\right)
\eeq
such that the partition function associated to our system is given by
the {\it Fermi-Dirac}
Distribution
 \begin{equation}\lb{pf}
 {\mathcal Z}=\prod_{\tau,\tau n, {\tilde p}, m_s} \left[1+e^{{\be(\tilde{\mu}-
E_{\tau, \tau n, {\tilde p}, m_s})}}\right]
\end{equation}
where ${\tilde{\mu}}$ is the chemical potential of particles,
$\tau$ takes plus one when  $n>0$ and minus one otherwise. $\be={1}{k_BT}$
with $K_B$ is Bolzaman constant and $T$ is the temperature.
We define the shorthand
notation $\{{\bf l}\}=\tau, \tau n, {\tilde p}, m_s$ to be used in the next.
Using~(\ref{pf}) to write \eqref{GTP1}
\begin{equation}
\Omega=-\frac{1} {\be}\sum_{\{{\bf l}\}}\ln\left[
1+e^{\beta({\tilde{\mu}-E_{n{\tilde p}s}})}\right].
\end{equation}

It is convenient to adopt
the dimensionless variable
$\mu={\tilde{\mu}\over mc^2},\
\varepsilon_{n{\tilde p}s}= {E_{n{\tilde p}s}\over mc^2}\
\textrm{and} \ \theta={1\over \be mc^2}$.
Requiring
$\Delta\tilde p=-c{E\over B}{\tilde p}$ and assuming that
$\arrowvert{\tilde p}\arrowvert\leq {eBL_y\over
  2c}$ is fulfilled, we write
GTP as
\begin{equation}
\Omega=-mc^2\theta N_{\phi} \int_{-b/2}^{b/2} \frac{d\tilde p}{b}\sum_{\{{\bf l}\}}\ln\left[
1+e^{{\mu\over  \theta}}\ e^{-{\varepsilon_{n{\tilde p}s} \over \theta}}\right].
\end{equation}
where $b=\frac{eBL_y}{mc^2}$ and $N_{\phi}= \frac{eBS}{hc}$ is
the number of quantum electron states in the magnetic field for a given $n$ in an area
$S=L_xL_y$.
To evaluate GTP, we use the Mellin
transformation method with respect to the variable
$e^{\mu\over \theta}$. After calculation, we obtain
\begin{equation}
\Omega=\mp 2\epsilon\te\sum_{s=-\infty}^{\infty}{\sf{Res}}\left [\sum_{\{\bf l\}} 
{\pi
e^{s\mu\over \theta}\over s\sin\left(\pi
s\right)}
e^{-s{\varepsilon_{\{{\bf l}\}}\over \theta}}\right]
\end{equation}
where the minus (plus) sign refers the closing sense of the counter to
the left (right) of the imaginary axis for $\mu>0$ ($\mu<0$).
Now we show
\begin{widetext}
\begin{eqnarray}
\Omega = \mp \epsilon\te N_{\phi}\sum_{s=-\infty}^{\infty}{\sf{Res}}
\left[{\pi e^{s{\kappa\over \theta}{z\over 2\pi}} \over
s\sin\left(\pi s\right)}\left\{
-1+2\sum_{(\tau^3n)=0}^{+\infty}\left(e^{-{s\over \te}\sqrt{2\kappa v_{F}^{2}}}\right)^{\sqrt{\tau^3n}}\right\}\right.
\left. \sum_{m_s=\pm {1\over 2}}\left(e^{-s{g^{\ast}\kappa\over\theta}}\right)^{m_s}
\int_{-b/2}^{b/2}e^{s{ec{\tilde p}\hbar E\over \epsilon^2\te\kappa}}{d{\tilde p}\over b}\right].
\end{eqnarray}
\end{widetext}
where $g^{\ast}={g\epsilon\mu_B\over e\hbar c^2}$,  $\kappa={ec\hbar B\over \epsilon^2}$ and $\epsilon=mc^2$.
After integration, we end up with
\begin{widetext}
\begin{equation}\lb{GTP2}
\Omega=\mp 2 \epsilon \te N_{\phi}\sum_s
{\sf{Res}}\left[{\pi
  e^{s{\kappa\over \theta}{z\over 2\pi}}\over s\sin\left(\pi s\right)}\coth\left({s\over \theta}\sqrt{\kappa
v_{F}^{2}\over2}\right)\cosh\left({sg^{\ast}\kappa\over2\theta}\right){\sinh\left(seEL_y/2\epsilon \te\right)\over
  seEL_y/2\epsilon \te}\right]
\end{equation}
\end{widetext}
where $z={2\pi\mu\over\kappa}$.
For the residue calculations, we distinguish two specials parts of
GTP $\Omega=\Omega_{\sf mon}+\Omega_{\sf mon}$. The first concerning the real poles called the monotonic
part $(\Omega_{\sf mon})$. But the second is related to the imaginary
poles called the oscillating part $(\Omega_{\sf osci})$. In our
analysis, we restrict the calculation of $\Omega_{\sf mon}$ and
$\Omega_{\sf osci}$ only for the minus sign, i.e for $\mu>0$.
We calculate $\Omega_{\sf mon}$ in $s=0$ and we neglect the
 contribution of other real poles. This gives
 \begin{widetext}
\begin{equation}
\Omega_{\sf mon}\approx -2\epsilon  N_{\phi}\beta\left[{1\over
    3}+{ {g^{\ast}}^2\over
8}\left({\kappa\over \lam}\right)^2+{z^2\over 8\pi^2}\left({\kappa\over \lam}\right)^2+{\al^2 \over 24}\left({\kappa\over \lam}\right)^2+{\al\pi^2\over
6}\left({\te\over \lam}\right)^2\right].
\end{equation}
\end{widetext}
Let us evaluate $\Omega_{\sf osci}$ in the poles $s_l={i\pi l\theta\over \lam}$ with
$l=1, 2, 3,\cdots$. Indeed, at low temperature and strong magnetic field, i.e $\te\ll\kappa$, we
obtain
\begin{widetext}
\begin{equation}
\Omega_{\sf osci}\approx-4 \epsilon N_{\phi}\lam\sum_{l=1}{(-1)^{l+1}\over\pi^2l^2}\cos\left({z\kappa\over
2\lam}l\right)\cos\left({\kappa g^{\ast}\pi\over \lam}l\right)
  {\sin\left({\al\pi\kappa/2\lam}\right)\over{\al\pi\kappa/ 2\lam}}
\end{equation}
\end{widetext}
where $\al={eEL_y\over\kappa \epsilon}$ and $\lam=\sqrt{\kappa
  v_{F}^{2}\over 2}$. Now combining all and using the assumption of very weak electric field ($\al\ll 1$)
  to write \eqref{GTP2} as
\begin{equation}\lb{omi}
\Omega\approx- \epsilon N_{\phi}\lam\left[{2\over
    3}+ \left({g^{\ast}\kappa\over 2\lam}\right)^2+\left({z\kappa\over \pi\lam}\right)^2 +{4\over \pi^2}\Gamma(z)\right]
\end{equation}
where $\Gamma(z)$ is a periodic function
of ${z\kappa\over
2\lam} $ defined as
\beq\lb{ga}
\Gamma(z)=\sum_{l=1}{(-1)^{l+1}\over\l^2}\cos\left({z\kappa\over
2\lam}l\right)\cos\left({\kappa g^{\ast}\pi\over \lam}l\right).
\eeq
In what follows,
the above function will play a crucial role in getting the quantized Hall
plateaux for {\it Dirac} particles in graphene. 

\section{Hall conductivity}

To evaluate the Hall conductivity, we determine the
number of charge carriers responsible for conduction in our system
through the relation
\begin{equation}
N=-{1\over \epsilon}{\partial\Omega\over \partial\mu}.
\end{equation}
Thus according to~(\ref{omi}), $N$ can be easily derived as
\begin{equation}\lb{n}
N={N_{\phi}\over\pi}\left( {\kappa\over \lam}\right)
\left[ z+8\left({\lam\over\kappa}\right)^2{d\Gamma(z)\over dz}\right].
\end{equation}
To proceed further, let us write (\ref{ga})  as
\begin{widetext}
\beq\lb{rol}
\Gamma(z)={1\over 2}\sum_{l=1}^{\infty}{(-1)^{l+1}\over
  l^2}\left\{\cos\left(\left[{k\over \lam}\left({z\over
  2}+g^{\ast}\pi\right)-2\pi\right]l\right)
+\cos\left({k\over \lam}\left({z\over
  2}-g^{\ast}\pi\right)l\right)\right\}
\eeq
\end{widetext}
and using the relation
\cite{grad}
\beq\lb{rl}
\sum_{l=1}^{\infty}{(-1)^{l+1}\over
  l^2} \cos(lx)={\pi^2\over 12}-{x^2\over 4},\qquad -\pi\leq x\leq\pi
\eeq
to show the result
\begin{widetext}
\beq\lb {gam}
\Gamma(z)=\left\{ \begin{array}{ll}
-{5\pi^2\over12} -  {1\over 4}\left({\kappa\over \lam}\right)^2
\left({z^2\over 4}+ {
    g^\ast}^2\pi^2\right) +{\pi\over 2}\left({\kappa\over
  \lam}\right)\left({z\over 2}+ g^\ast\pi\right)
  \qquad \qquad \qquad  \qquad\qquad\qquad\ \ \textrm{if }\
z\in{\mathbf{ I_1}}\cap {\mathbf{ I_2}}
\\
{\pi^2\over12} -  {1\over 4}\left({\kappa\over \lam}\right)^2\left({z^2\over 4}+ {
    g^\ast}^2\pi^2\right) -
\left({i^2+\left(i+2\right)^2\over 8}+{1\over 2}\left({\kappa\over
  \lam}\right)g^\ast\right)\pi^2+{1\over 4}\left(i+1\right)\left({\kappa\over
  \lam}\right)\pi z
 \qquad \textrm{if }\
z\in{\mathbf{ I_3}}\cap {\mathbf{ I_4}}
\end{array} \right.
\eeq
\end{widetext}
where $i$ is an even integer and ${\mathbf{ I_j}},\ j=1, 2, 3, 4$ are intervals defined as
\newpage
\begin{widetext}
\bqr
&&{\mathbf{I_1}}=\left[2\left({\lam\over\kappa}-g^\ast\right)\pi, 2\left({3\lam\over\kappa}-g^\ast\right)\pi \right], \qquad
{\mathbf{I_3}}=\left[2\left({\beta\over
    \kappa}\left(i+1\right)-g^\ast\right)\pi, 2\left({\lam\over
    \kappa}\left(i+3\right)-g^\ast\right)\pi \right] \nonumber\\
&&{\mathbf{I_2}} =  \left[2\left( g^\ast-{\lam\over
    \kappa}\right)\pi, 2\left( g^\ast+{\lam\over
    \kappa}\right)\pi \right], \qquad
    {\mathbf{I_4}}=
\left[2\left({\lam\over
    \kappa}\left(i-1\right)+g^\ast\right)\pi,2\left({\lam\over
    \kappa}\left(i+1\right)+g^\ast\right)\pi \right]\nonumber
    \eqr
\end{widetext}

To describe the quantum Hall effect, it is essential to evaluate
the Hall conductivity $\sigma_{\sf H}$. Hence, using the Drude model, $\sigma_{\sf H}$ is
\beq\lb{hc}
\sigma_{\sf H}=-{\rho c e\over B}
\eeq
where $\rho$ is the particle number per unit area. In function of
the degree of the degeneracy of each LL $ N_{\phi}$ and the particle
number, $\sigma_{\sf H}$ is expressed in terms  of {\it
  von Klitzing} conductance ${e^2\over h}$, as
\beq\lb{hc1}
\sigma_{\sf H}=-{N\over N_{\phi}}{e^2\over h}=-\nu{e^2\over h}
\eeq
where $\nu$ is the filling factor of LL.
Using (\ref{n}), to obtain
\beq\lb{hc2}
\sigma_{\sf H}=-{1\over\pi}\left( {\kappa\over \lam}\right)\left[ z+8\left({\lam\over\kappa}\right)^2{d\Gamma(z)\over dz}\right]{e^2\over h}.
\eeq
This compared to \eqref{hc1} gives
\beq\lb{ff}
\nu={1\over\pi}\left( {\kappa\over \lam}\right)
\left[ z+8\left({\lam\over\kappa}\right)^2{d\Gamma(z)\over dz}\right].
\eeq
Now using (\ref{gam}) to find 
\beq
\nu=\left\{ \begin{array}{ll}
2
  \qquad \qquad \qquad  \qquad\textrm{if }\
z\in{\mathbf{ I_1}}\cap {\mathbf{ I_2}}
\\
2\left(i+1\right)
  \qquad  \qquad \  \textrm{if }\
z\in{\mathbf{ I_3}}\cap {\mathbf{ I_4}}
\end{array} \right.
\eeq
Recall that $i$ is taking even value and then we can write $i=2n$
to recover the famous result $\nu=4\left(n+\frac{1}{2}\right)$, with
$n$ is an integer.
This clearly shows how one can describe
the integer quantum Hall effect in graphene
based on the thermodynamical properties
of our system.


\section{Conclusion}

By taking into account of the edges effects  in terms of a pseudo-potential in
a monolayer graphene, we
have shown that the corresponding Hall conductivity undergoes to a sequence of
plateaux. Its quantization is performed by make using 
of the {\it Fermi-Dirac} statistical.
This has been done by considering a system of Dirac fermions in graphene
submitted to an electromagnetic field and evaluating
 the grand thermodynamical potential as well as related physical quantities
 like number of particles.

\section{Acknowledgment}

The generous support provided by the Saudi Center for Theoretical Physics (SCTP)
is highly appreciated by the author.

\end{document}